\documentclass{rspublic}
\usepackage{amsmath}
\usepackage{amsfonts}
\usepackage{amssymb}
\usepackage[Symbol]{upgreek}
\usepackage[nointegrals]{wasysym}
\usepackage{mathrsfs}
\usepackage{graphicx}
\newcommand{\real}{{\mathbb R}}
\newcommand{\alr}{\alpha r}
\newcommand{\alt}{\alpha t}

\newcommand{\zt}{z_0 t^2}
\newcommand{\zr}{z_0 r^2}
\newcommand{\psix}{\psi_{x}}
\newcommand{\psir}{\psi_{r}}
\newcommand{\psit}{\psi_{\theta}}
\newcommand{\psione}{\psi_{1}}
\newcommand{\psitwo}{\psi_{2}}
\newcommand{\phione}{\phi_{1}}
\newcommand{\phitwo}{\phi_{2}}
\newcommand{\psirone}{{\psi_{r_1}}}
\newcommand{\psirtwo}{{\psi_{r_2}}}
\newcommand{\psirthree}{{\psi_{r_3}}}
\newcommand{\psirfour}{{\psi_{r_4}}}
\newcommand{\psirfive}{{\psi_{r_5}}}
\newcommand{\psirsix}{{\psi_{r_6}}}
\newcommand{\proneone}{{\psi'_{r_1}}}

\newcommand{\prthreeone}{{\psi'_{r_3}}}
\newcommand{\prfourone}{{\psi'_{r_4}}}
\newcommand{\prfiveone}{{\psi'_{r_5}}}
\newcommand{\prsixone}{{\psi'_{r_6}}}

\newcommand{\psitone}{{\psi_{\theta_1}}}
\newcommand{\psittwo}{{\psi_{\theta_2}}}
\newcommand{\psitthree}{{\psi_{\theta_3}}}
\newcommand{\psitfour}{{\psi_{\theta_4}}}
\newcommand{\psitfive}{{\psi_{\theta_5}}}
\newcommand{\psitsix}{{\psi_{\theta_6}}}
\newcommand{\ptoneone}{{\psi'_{\theta_1}}}
\newcommand{\pttwoone}{{\psi'_{\theta_2}}}
\newcommand{\ptthreeone}{{\psi'_{\theta_3}}}
\newcommand{\ptfourone}{{\psi'_{\theta_4}}}
\newcommand{\ptfiveone}{{\psi'_{\theta_5}}}
\newcommand{\ptsixone}{{\psi'_{\theta_6}}}

\newcommand{\psixone}{{\psi_{x_1}}}
\newcommand{\psixtwo}{{\psi_{x_2}}}
\newcommand{\psixthree}{{\psi_{x_3}}}
\newcommand{\psixfour}{{\psi_{x_4}}}
\newcommand{\psixfive}{{\psi_{x_5}}}
\newcommand{\psixsix}{{\psi_{x_6}}}
\newcommand{\pxoneone}{{\psi'_{x_1}}}
\newcommand{\pxtwoone}{{\psi'_{x_2}}}
\newcommand{\pxthreeone}{{\psi'_{x_3}}}
\newcommand{\pxfourone}{{\psi'_{x_4}}}
\newcommand{\pxfiveone}{{\psi'_{x_5}}}
\newcommand{\pxsixone}{{\psi'_{x_6}}}
\newcommand{\pxonetwo}{{\psi''_{x_1}}}
\newcommand{\pxtwotwo}{{\psi''_{x_2}}}
\newcommand{\pxthreetwo}{{\psi''_{x_3}}}
\newcommand{\pxfourtwo}{{\psi''_{x_4}}}
\newcommand{\pxfivetwo}{{\psi''_{x_5}}}
\newcommand{\pxsixtwo}{{\psi''_{x_6}}}
\providecommand{\abs}[1]{\lvert\,#1\,\rvert}
\newcommand{\gfrac}[2]{\genfrac{}{}{0pt}{0}{#1}{#2}}
\begin{document}
\title[Pipe Poiseuille flow]{\Large{Short-time evolution of pipe Poiseuille flow}}
\author[F. Lam]{F. Lam}
%
%
\label{firstpage}
\maketitle
\begin{abstract}{Navier-Stokes Equations; Pipe Poiseuille flow; Linearization; Viscosity: Diffusion; Eigen-Value Spectrum}
In the present paper we prove that the pipe Poiseuille flow of parabolic velocity profile attenuates exponentially in time with respect to three dimensional infinitesimal disturbances of the form
\begin{equation*}
 \exp\big( \: \ri \alpha(x-ct)+ \ri n\theta \: \big)
\end{equation*}
at all finite wave numbers $\alpha$ and Reynolds numbers for given azimuthal periodicity $n \geq 0$ if the equations of motion are linearized. The spectra of the eigenvalue $c$ are shown to consist of infinitely many discrete eigen-modes. Results of asymptotic analysis, expressed in simple algebraic formulas and functional relations, are given. Good comparison has been found in the approximations and numerical computations. The present results are best interpreted as a description of the pipe flow regime, where the linear diffusion due to viscosity dominates.  
\end{abstract}
\section{Introduction}
The present paper deals with the linearised Poiseuille flow in a straight circular pipe subject to three-dimensional infinitesimal disturbances. The fluid is treated as a viscous, incompressible fluid and is assumed to be uniform throughout the pipe. Experiments have demonstrated that both laminar and turbulent motions exist, as first shown in the classic experiment of Reynolds (1883). The conventional wisdom ascribes breakdown of the laminar flow to the instability caused by wave-form disturbances (see, for example, Landau \& Lifshitz 1987). 
A series of theoretical attempts have been made, over a time span of more than a century, to explain the observed flow phenomena by considering the linear stability of the flow: Sexl 1927$a$, 1927$b$; Sexl \& Spielberg 1930; Synge 1938; Pretsch 1941; Pekeris 1948; Corcos \& Sellars 1959; Lessen, Sadler \& Liu l964; Gill 1965; Lessen, Sadler \& Liu l968; Burridge \& Drazin 1969; Drazin \& Davey 1972; Garg \& Rouleau 1972; Salwen \& Grosch 1972; Salwen, Cotton \& Grosch 1980; Wadih 1984; Meseguer \& Trefethen 2003; Walton 2004. For the connection of the pipe Poiseuille flow to other stability theories, we refer to the reviews by Lin (1955), Stuart (1966), Drazin \& Reid (1981) and Schmid \& Henningson (2001). 

Within the framework of the continuum, fluid mechanics is governed by the Navier-Stokes equations. The equations of motion are a set of partial differential equations whose solutions must be sought as an {\it initial-boundary} value problem. For a rigid non-permeable solid pipe, boundary condition is the no-slip condition on the wall. In practice, appropriate initial conditions must be supplied in order that the equations of motion admit unique solution. Because of the analytical difficulties in dealing with the equations of motion, various approximation schemes have been proposed. Notably, an {\it assumption} has been put forward that the equations of motion may be linearized (see, for example, p1 of Lin 1955). By considering wave-like disturbances of given wave-numbers, an eigen-value problem is obtained, and a dispersion law determines whether the perturbation waves grow or attenuate in time for specific Reynolds number. This is known as the linear stability analysis. For a given experimental set-up, a fully-developed laminar flow approaches a parabolic velocity profile at {\it one location} downstream of the pipe inlet. Investigations suggested that the inlet flow must be carefully controlled in order to achieve the parabolic profile. In other words, the location depends on the initial condition. The laminar-turbulent transition will inevitably occur downstream and becomes a function of the inlet flow. Theoretically, we consider that the flow in the pipe as a solution of the {\it complete} Navier-Stokes equations with specified initial conditions. If disturbances are present or introduced into the flow at time $t_m$, (while keeping the boundary condition), the modified flow from time $t_m$ onward is described by the equations with modified initial data. Since the initial data have fully taken into account any disturbance; the effect of disturbing a flow must be examined by solving the complete equations of motion. Dissimilar initial conditions (with the no-slip condition on the pipe wall) correspond to the variation in the Reynolds number. Thus the laminar flow at any fixed location, assuming existed, does not necessarily possess an identical velocity profile; the parabolic distribution does not exist at Reynolds number of arbitrary magnitude. The stability concept of a fixed mean flow as a function of Reynolds number becomes difficult to justify. This is because the complete equations do not admit similarity solutions except over a small time interval from the start of a motion when the equations of motion govern linear diffusion. Evidently, the interval depends on the initial data. Moreover, it has been shown that, as an {\it a priori} bound, the total vorticity is an invariance of fluid motion in $\real^3$. This is an intrinsic property of the incompressible Navies-Stokes equations and its implications in flow evolution are of significance (Lam 2013). By analogy, the development of vorticity in pipe flow must be self-regulated accordingly if we consider initial flows of finite-energy. A disturbance prescribed in the linear stability analysis would introduce spurious vorticity into the flow-field in all likelihood since the dispersion law is established independent of the invariance or of the initial conditions. 

Nevertheless, if we are interested in the development of small initial data over a small time interval such that the viscous diffusion dominates, then the use of the linearized equations may be an acceptable approximation. In other words, we may tabulate a data-base of the normal modes of every Fourier component as long as the eigen-modes are all damped. Evidently, in these circumstances, the approximate solutions could be used to describe a ``mean" stage of the laminar flow in a pipe as time $t \rightarrow 0$, though our theory can never provide an answer to the puzzle of turbulence initiation. Technically, a laminar flow always exists and is a solution of the full equations of motion in a brief period of time from the commencement of any fluid motion. 

In this paper we show that all the eigen-modes of infinitesimal disturbances in the pipe Poiseuille flow are damped for all Reynolds numbers and wave numbers. Some properties of the eigenvalues spectra are discussed. Relevant asymptotic theory is also developed. We then examine the damped modes in the limit of vanishing Reynolds number, $R \rightarrow 0$. Lastly, we discuss the related issues of purely two-dimensional disturbances.
\section{Problem formulation}
We consider a Newtonian fluid with density $\rho$ and viscosity $\mu$ in a circular pipe of diameter $d$. All dimensional quantities in the physical problem are made dimensionless by using the length and velocity scales $d$ and $V_{0}$, the maximum velocity at the pipe centre. The time and pressure are normalized by $d/V_0$ and $\rho V_0^2$ respectively. In a cylindrical co-ordinate system $(x, r, \theta)$, the basic velocity profile is given by
\begin{equation} \label{prof}
V(r)=1-r^{2}.
\end{equation} 
The Reynolds number is
\begin{equation*}
R=\rho V_{0}d/\mu.
\end{equation*}
Let the perturbation velocity be 
\begin{equation*}
u=(u_{x}, u_{r}, u_{\theta}).
\end{equation*}
The linearized equation of motion (Batchelor \& Gill 1962) takes the form of
\begin{equation} \label{linear-ns}
{\partial u} / {\partial t} + (1-r^2){\partial u }/{\partial x}-2r u_{r}
=-{\nabla{P}}/{\rho}+{\nu}{\nabla}^2 u,
\end{equation}
where $P$ is the pressure perturbation. The equation of continuity reads
\begin{equation} \label{continuity}
\nabla.u = {\partial{u_{x}}}/{\partial{x}}+{\partial{(ru_{r})}}/(r{\partial{r}})+
{\partial{u_{\theta}}}/(r{\partial{\theta}})=0.
\end{equation}
The velocity and the pressure have their Fourier components,
\begin{equation} \label{dists}
u_{x},u_{r},u_{\theta},P/ \rho=\Real \Big( \: \big( \psi_{x}(r),\ri {\psi}_{r}(r),\psi_{\theta}(r), p(r) \big)
\: \exp \big( \: \ri {\alpha}(x-ct)+\ri n{\theta} \: \big)\: \Big),
\end{equation}
where $\Real$ denotes that the real part of the quantity in the brackets. The disturbance waves are described by wave number $\alpha$ and azimuthal periodicity $n$. For given Reynolds, we seek to establish the growth or decay of the waves by considering the eigenvalue 
\begin{equation*}
c=c_{r}+\ri c_{i},
\end{equation*}
where $c_{r}$ is the phase speed of the disturbance. If the imaginary part 
$c_{i}$ turns out to be negative, then the disturbance will attenuate like
\begin{equation*}
u \propto \exp(- |c_i| t)
\end{equation*}
over a short time interval. Hence the disturbed flow cannot substantially deviate from the local steady profile in the presence of imposed disturbances. Superposition of all the Fourier components may provide an approximate solution for flows whose initial data closely resemble the parabolic distribution.

On substituting (\ref{dists}) in (\ref{linear-ns}) and (\ref{continuity}) we obtain a system of four ordinary differential 
equations for the unknowns $\psi_{x}, \psi_{r}, \psi_{\theta}$ and $p\:$:
\begin{equation} \label {u-comps}
\begin{split}
\psix''+\frac{\psix'}{r}
-\left( \frac{\alpha^2+n^2}{r^2}\right) \psix -\ri \alpha R \big(1-r^2-c\big) \psix+2 \ri Rr \psir & =\ri {\alpha}R p, \\
\psir''+\frac{\psir'}{r}
-\left( \frac{\alpha^2+(n^{2}+1)}{r^2}\right) \psir - \ri \alpha R \big(1-r^2-c\big){\psir}- \frac{2n \psit}{r^2} & =-\ri Rp', \\
\psit''+\frac{\psit'}{r}
-\left( \frac{\alpha^2+(n^{2}+1)}{r^2}\right) \psit - \ri \alpha R \big(1-r^2-c\big) \psit- \frac{2n \psir}{r^2} & =\ri Rnp/r, \\
\alpha \psix+ \psir'+ \frac{\psir}{r}+ \frac{n \psit}{r} & =0,
\end{split}
\end{equation}
where a prime denotes differentiation with respect to $r$. The boundary conditions to be satisfied at the pipe wall are simply the no-slip condition
\begin{equation} \label{wall-bc}
\psix(1)=\psir(1)=\psit(1)=0\;\;\;\;{\mbox{for all}}\;\;n. 
\end{equation}
From physics point of view, we also require that $\psir,\;\psit,\;\psix$ and $p$ are bounded throughout the pipe.
Fourier transform of the equations of the motion with respect to $\theta$
requires $n$ in (\ref{dists}) to take integer values, $0,{\pm}1,{\pm}2,\ldots$, so as to ensure the transformed functions to be singled-valued.
We notice that the set of the equations is invariant if
$\theta$ is replaced by $-\theta$ and $\psit$ by $-\psit$.

Multiplying the first equation of (\ref{u-comps}) by $\alpha$ and the third by $n/r$, dividing the second by $r$ and differentiating the second with respect to $r$, 
the resulting four equations can be combined, using the fourth of (\ref{u-comps}) which is the continuity, to
\begin{equation} \label {eq:pressure}
p''+p'/r-\left(\alpha^2+n^2/r^2 \right)p=2{\alpha}V' \psir = -4{\alpha}r \psir.
\end{equation}
This equation states that the disturbance pressure in any plane normal to the axis of the pipe 
is solely driven by the radial velocity component in uni-directional mean flows such as (\ref{prof}). 
The equations of motion can also be transformed into the orthogonal co-ordinate system of Batchelor \& Gill (1962), in which the disturbances can be described in components being parallel and normal to the local helices of constant phase.
The radial component is however invariant in this transformation. 
\section{Analysis of the case $n=0$}
\subsection{Torsional modes}
For rotationally symmetric disturbances proportional to $\exp \big(\ri {\alpha}(x-ct)\big)$, the third equation of (\ref{u-comps}) becomes
\begin{equation} \label{theta0}
	L_1 \psit - \ri \alpha R(1-r^2-c) \psit=0
\end{equation}
where $D=\rd/\rd r$, and 
\begin{equation*}
L_1=D^2 + D/r -(\alpha^2 + 1/r^2).
\end{equation*}
Set $\phi=r\psit$. 
By a change of the independent variable, 
\begin{equation*}
z=( \alpha R)^{1/2}\exp(\ri 3 \pi /4)\:r^2 = z_{0}\:r^{2},\;\;\;\;\;\Real(z) <0, 
\end{equation*}
equation (\ref{theta0}) can be transformed into a Whittaker equation:
\begin{equation} \label{whit0}
\phi''(z)+\left(-1/4+\kappa/z \right) \phi(z)=0,
\end{equation}
where
\begin{equation} \label{kap}
\kappa=-\frac{\exp({-{\pi}\ri /4})}{4} \Big( (1-c)(\alpha R)^{1/2} -\ri \: {\alpha^2} {(\alpha R)^{-1/2}} \Big) = 1-a.
\end{equation}
Two linearly independent solutions of (\ref{whit0}) are readily found and expressed in terms of the confluent hypergeometric functions (see, for example, Abramowitz \& Stegun 1972; Lebedev 1972). Thus the solutions of (\ref{theta0}) are written as
\begin{equation}
\begin{split}
	\psione(r) & = z_0 r \re ^{-z_0 r^2 /2}M(a,2,z_0 r^2),\\
	\psitwo(r) & = z_0 r \re ^{-z_0 r^2 /2}U(a,2,z_0 r^2).
\end{split}
\end{equation}
Denote $h_0$ by Green's function for (\ref{theta0}) satisfying the boundary conditions. We find 
\begin{equation*} 
h_{0}(r,s,c)=-\frac{\ri \: \Gamma (a)}{2 z_0 \psione(1) s }
\begin{cases}
\:{\psione}(r)\:\Big({\psione}(1){\psitwo}(s)
-{\psione}(s){\psitwo}(1) \Big), & 
\;\;\;\;\;0{\leq}r{\leq}s, \\
 & \\
\:{\psione}(s)\:\Big( {\psione}(1){\psitwo}(r)
-{\psione}(r){\psitwo}(1) \Big), & 
\;\;\;\;\;s{\leq}r{\leq}1, \\
\end{cases}
\end{equation*}
where $\Gamma$ is the gamma function.
The singularity of Green's function is defined by the condition $\psi_1(1)=0$. Thus
the dispersion relation is given by
\begin{equation} \label{t-modes}
 \Delta_{T}(\alpha,R,c)=z_0 \exp({- z_0 / 2}) M(a,2,z_{0})=0.
\end{equation}
These eigen-modes, known as the torsional modes, coincide with the zeros of Kummer's function. Referring to appendix A, function $M(a,2,z_0)$ has no complex zeros if $c_i > -\alpha/R$; all the modes defined by (\ref{t-modes}) are damped. Synge (1938) and Pekeris (1948) have established the stability from a different point of view.

Given Green's function $h_0$, it is straightforward to reduce (\ref{theta0}) to a homogeneous Fredholm integral equation of the second kind:
\begin{equation} \label{t-fred}
	\psi_{\theta}(r) =- \lambda\int_{0}^{1} H_{0}(r,s)\: \psi_{\theta}(s) \:\rd s,
\end{equation}
where $\lambda=-\ri \alpha R c$, the kernel $H_0 (r,s) = h_0 (r,s,c=0)$, and $H_0(0,s) = H_0(1,s) = 0$, for $s > 0$. It is Hermitian $H_0(r,s)=H_0^*(s,r)$. Since function $M(a,2,z_0 r^2) \rightarrow 1$ as $r \rightarrow 0$, $H_0(0,s)$ is bounded as $s \rightarrow 0$. Moreover, $H_0(r,s)$ is everywhere continuous for $0 \leq r,s \leq 1$, and $r\neq s$. The jump in $H_0$ at $r{=}s$ is 
\begin{equation*}
	{ \partial (rH_0) }/{\partial s} \Big|_{s=r+0}-{ \partial (rH_0) }/{\partial s} \Big|_{s=r-0}=-1,
\end{equation*}
which is necessarily finite except at the origin. Since $H_0(0,s){=}0$, $H_0$ satisfies the Lipschitz condition 
\begin{equation*}
	\abs{ H_0(r,s_{1}) - H_0(r,s_{2}) }\;{\leq}\;M_0\; \abs{ s_{1}-s_{2} }
\end{equation*}
for some constant $M_0$ for $0 \leq r,s_1,s_2 \leq 1$. For finite $\alpha$ and $R$, $\abs{H_0}$ must be bounded, that is, 
\begin{equation*}
\abs{H_0} < N_0
\end{equation*}
for some constant $N_0$ from the asymptotic properties of hypergeometric functions function (\ref{eq:m-asymp}) and (\ref{eq:u-asymp}). Then the integral in the Fredholm equation (\ref{t-fred}) can be approximated by finite difference, and thus the equation is equivalent to an algebraic system. According to theory of Fredholm integral equation (see, for example, Tricomi 1957), the eigen-value relation (\ref{t-modes}) has an alternative form, known as the Fredholm determinant,
\begin{equation*}
\Delta_{T}(\lambda) = \sum_{k=0}^{\infty} \frac{(-\lambda)^k}{k!} \int_0^1 \int_0^1 \cdots \int_0^1 F(H_0) \rd r_1 \rd r_2 \cdots \rd r_k,
\end{equation*}
where $F(H_0)$ is the Fredholm kernel, and $F(H_0) \leq (N_0 \sqrt{k})^k$. Thus the series for the eigen-value relation converges for all values of $\lambda$ as we may apply a ratio test on the majorant 
\begin{equation*}
 |\Delta_T(\lambda)| \leq \sum_{k=0}^{\infty} \big( |\lambda| (N_0 \sqrt{k}) \big)^k / k!.
\end{equation*}
In other words, $\Delta_T(c; \alpha,R)$ is an entire function of $c$ for given finite $\alpha$ and $R$. Evidently, $c=0$ is not an eigen-value. Since the kernel $H_0$ is Lipschitz continuous, the order of the entire function $\Delta_{T}(c)$ is at most $2/3$, in view of the theory for entire functions (see, for example, Boas 1954; Hochstadt 1973; Levin 1964). Dispersion $\Delta_T(c)$ can be expanded according to Hadamard's factorisation theorem:
\begin{equation*}
\Delta_T(c) = \exp\big(ac\big) \prod_{m=1}^{\infty} \Big( 1 - \frac{c}{c_m}\Big) \exp\Big(\frac{c}{c_m} \Big),
\end{equation*}
where $c_m$ are the elements of the eigen-value spectrum, and $a$ is a positive constant (which can be estimated according to the asymptotic expansion). Evidently, $H_0(r,s)$ cannot be written in a degenerated form of $\sum_{j=1}^{k} T_j(r) S_j(s)$ for some bounded functions $T(r)$ and $S(s)$. It follows that $\Delta_{T}(c)$ cannot be a polynomial in $c$, that is, it possesses infinitely many zeros. As the Hermitian kernel $H_0$ has at least one non-zero eigenvalue, 
we see that the spectrum of the dispersion relation $\Delta_T$ must be non-empty. In fact, it consists of denumerable numbers of the eigenvalues $c$ for finite $\alpha$ and $R$. {\it There exist infinitely many solutions of $\Delta_T$ for given wave-number $\alpha$ and Reynolds number $R$, and }$c_i \leq -\alpha/R$. 

The completeness of the eigen-value spectrum may be proved as a Sturm-Liouville eigen-value problem from theory of ordinary differential equations (see, for example, Chapter 7 of Coddington \& Levinson 1955).
\subsection{Meridional modes}
The second part of the eigenvalue spectrum for axial symmetric disturbances $n{=}0$ consists in the radial disturbance component. By the method of variation of parameters, we find that the solution for equation (\ref{eq:pressure}) can be written as
\begin{equation} \label{press0}
 p'(r)/\alpha = C_1 I_1(\alr) + C_2 K_1(\alr) + \int_0^{r} t^2 X_0(r,t) \psir(t) \rd t,
\end{equation}
where $C_1$, $C_2$ are constants, and
\begin{equation*}
 X_0(r,t)=-4 \alpha^2 \: \Big( I_1(\alr) K_0(\alt) + I_0(\alt) K_1(\alr) \Big).
\end{equation*}
As usual, the notations, $I_n$ and $K_n$, denote the modified Bessel functions of the first and second kinds, respectively. To get an expression for $\psir$, let $\phi_r=r \psir$. The second equation in (\ref{u-comps}) can be reduced to
\begin{equation} \label{whit0-phi}
 \phi''_r(z)+\Big(-\frac{1}{4}+\frac{\kappa}{z} \Big) \; \phi_r(z)=\frac{p'}{4 r \alpha}.
\end{equation}
The left-hand side is a Whittaker equation. Thus we obtain
\begin{equation*}
 \psir(r)=C_3 \psione(r) + C_4 \psitwo(r) + \frac{\Gamma(a)}{8 z_0} \int_0^r Z_0(r,t) t^{-2}p'(t)/\alpha \rd t,
\end{equation*}
where $C_3$ and $C_4$ are constant, and 
\begin{equation*}
Z_0(r,t) = -\Big( \psione(r)\psitwo(t)-\psione(t)\psitwo(r) \Big).
\end{equation*}
Combining with (\ref{press0}), $\psir$ satisfies the following Volterra integral equation of the second kind
\begin{equation} \label{phir0-volt}
  \psir(r) = \sum_{k=1}^{4} C_k f_k(r) + z_0^{-1} \int_0^r k(r,t) \psir(t) \rd t,
\end{equation}
where the kernel is given by
\begin{equation*}
  k(r,t) = \frac{\alpha \Gamma(a) } {2}  \int_t^r Z_0(r,s) X_0(s,t) \rd s,  
\end{equation*}
and the functions $f_k$ are given by 
\begin{equation*}
	\genfrac{}{}{0pt}{}{f_1(r)}{f_2(r)}= \genfrac{}{}{0pt}{}{\psione(r), }{ \psitwo(r), }\;\;\;\;\;\;
	\genfrac{}{}{0pt}{}{f_3(r)}{f_4(r)}= \frac{\Gamma(a)}{8 z_0} \int_{0}^r Z_0(r,t) \: \genfrac{}{}{0pt}{}{I_1(\alt) \rd t, }{ K_1(\alt) \rd t. }
\end{equation*}
Explicitly, the solution of $\psir$ is given by
\begin{equation} \label{phir0-volt-sol}
\psir(r)  = \sum_{k=1}^{4} C_k {\psir}_k(r)=\sum_{k=1}^{4} C_k \Big( \: f_k(r) + z_0^{-1} \int_{0}^{r} h(r,t) f_k(t) \rd t \: \Big). \\
\end{equation}
The resolvent kernel $h$ is related to the iterated kernel $k_j$ by
\begin{equation} \label{iter}
 \begin{split}
	h(r,t) & =\sum_{j=0}^{\infty} z_0^{-j} k_{j+1}(z,s),\\
	k_1 (r,t) & = k(r,t), \\
	k_{j+1} (r,t)& =\int_{t}^{r} k_l (r,s) k_m (s,t) \rd s,\;\;\;\;l=1,2,\dots j;\;m=j-l+1.
 \end{split}
\end{equation}
From the fourth equation of (\ref{u-comps}), that is, the continuity constraint, we see that the boundary condition $\psix(1)$ implies $\psir'(1)$. For flows of bounded velocity and vorticity, we require that $C_2=C_4=0$. Application of the two boundary conditions at the wall (\ref{wall-bc}) yields the eigenvalue relation
\begin{equation} \label{mm-modes}
\Delta_{M}(\alpha,R,c)= \left| \begin{array}{cc}
a_{11} & a_{12} \\
a_{21} & a_{22}  \\
\end{array} 
\right|
=a_{11}a_{22}-a_{12}a_{21}=0, 
\end{equation}
where
\begin{equation*}
	\genfrac{}{}{0pt}{}{a_{11}}{a_{12}}= \genfrac{}{}{0pt}{}{\psione(1)}{f_3(1)}+ \int_0^1  h(1,t) \: \genfrac{}{}{0pt}{}{\psione(t) \rd t, }{ f_3(t) \rd t, }
\end{equation*}
and
\begin{equation*}
	\genfrac{}{}{0pt}{}{a_{21}}{a_{22}}= \genfrac{}{}{0pt}{}{\psi'_1(1)}{f'_3(1)}+ \int_0^1  h'(1,t) \: \genfrac{}{}{0pt}{}{\psione(t) \rd t, }{ f_3(t) \rd t },
\end{equation*}
where $h'(1,t)=\partial h(r,t)/\partial r|_{r{=}1}$. 

Since the two rows in (\ref{mm-modes}) are two linearly independent boundary conditions at the wall, these rows cannot be in proportion. Nor can the sum or the difference of these two rows be in proportion. Similarly, ${\psir}_1$ and ${\psir}_3$ are the linearly independent solutions, the two columns (or their sum or their difference) cannot be in proportion.
Second, when $c_i > -\alpha/R$, none of $\psione(1)$, $\psione'(1)$, $\psitwo(1)$ and $\psitwo'(1)$ can vanish according to the results derived in appendix A. This in turn implies that $a_{11}$ cannot be zero. To see this, suppose that $a_{11}=0$, we must have $h(1,t) \psione(t) = - \psione'(t)$, from the fundamental theorems of calculus. But $h(1,1) \equiv 0$ while $\psione'(1) \neq 0$. Hence we have a contradiction. 
Next, if $a_{12} = 0$, it follows that $f'_3(1) = 0$. Then $a_{22}$ cannot vanish and hence the dispersion $\Delta_{M} \neq 0$. If $a_{12} \neq 0$, either $a_{21}$ or $a_{22}$ may be zero. It follows that $\Delta_{M} \neq 0$. In summary, {\it we have shown that $\Delta_{M}$ has no solutions for} $c_i > -\alpha/R$.
\subsection*{Alternative derivation of dispersion law}
To establish that $\Delta_M$ admits solutions for $c_i \leq -\alpha/R$, and to investigate the properties of the spectrum, it is convenient to proceed in an alternative way. Eliminating $p$ between the first and the second of (\ref{u-comps}), using the fourth (the continuity), we arrive at 
\begin{equation} \label{psi0}
	L_2(rL_1)\psir(r)=0,
\end{equation}
where 
\begin{equation*}
L_2 = D^2 - D/r - \big( \: \alpha^2 + \ri \alpha R (1-r^2-c) \: \big).
\end{equation*}
The boundary conditions are
\begin{equation} \label{psi0-wall-bc}
 \psir(1)=\psir'(1)=0. 
\end{equation}
Four linearly independent solutions of (\ref{psi0}) are readily found:
\begin{equation*}
 \begin{split}
	\psirone (r) & =  I_1(\alr),\\
	\psirtwo (r) & =  K_1(\alr), \\
	\psirthree (r) & =  \int_0^r Y_0(r,t) \zt \re ^{-\zt/2}M(a,2,\zt) \rd t=\phione(r),\\
	\psirfour (r) & =  \int_0^r Y_0(r,t) \zt \re ^{-\zt/2}U(a,2,\zt) \rd t=\phitwo(r), 
 \end{split}
\end{equation*}
where
\begin{equation*}
  Y_0= \alpha \Big( I_{1}(\alr)K_{1}(\alt)-I_{1}(\alt)K_{1}(\alr) \Big). 
\end{equation*}
Green's function associated with the fourth order system of (\ref{psi0}) and (\ref{psi0-wall-bc}) can be expressed in terms of these solutions as
\begin{equation} \label{eq:greens-g0}
g_{0}(r,t,c)=
\begin{cases}
 \;\overset{4}{\underset{j=1}\sum}  \;A_{j}(t) \;\psi_{r_j}(r), & 
\;\;\;\;\;0\;{\leq}\;t\;{\leq}\;r, \\
 \;\overset{4}{\underset{j=1}\sum}  \; B_{j}(t) \;\psi_{r_j}(r), & 
\;\;\;\;\;r\;{\leq}\;t\;{\leq}\;1, \\
\end{cases}
\end{equation}
where $A$'s and $B$'s are functions to be determined. For flows of bounded velocity and vorticity, we require that $A_2 = A_4 = 0$. The other unknown functions of $t$ are determined from the wall boundary conditions (\ref{psi0-wall-bc}), three continuity conditions for $g_0$, $g_0'$, $g_0''$ at $t=r$. We also have a jump in $g_0'''$ at $t=r$. In matrix notation, we have $A X = B$,
where $A=\{A_1 \;A_3\;B_1\;B_2\;B_3\;B_4 \}^T$, $B=\{0 \;0\;0\;0\;0\;{-1/t^3} \}^T$ (the superscript $T$ on the row vectors indicates the transpose).
We find that the determinant is given by
\begin{equation*}
\abs{A} = Q_0(\alpha) W_0 \Delta_{M}(\alpha,R,c),
\end{equation*}
where $W_0$ denotes the Wronskian of the linearly independent solutions, and the coefficient function $Q_0$ is constant for given $\alpha$,
\begin{equation*}
Q_0(\alpha)= \alpha^2 \Big( \; I_1(\alpha) \big(K_0(\alpha)+K_2(\alpha) \big) + K_1(\alpha) \big(I_0(\alpha)+I_2(\alpha) \big) \; \Big)/2.
\end{equation*}
The singularity of Green's function $g_0$ gives rise to the dispersion relation
\begin{equation} \label{m-modes}
	\Delta_M(\alpha,R,c)=\int_0^1 I_1(\alr) \:\zr \: \exp( {-\zr /2}) \: M(a,2,\zr) \rd r=0.
\end{equation}

To make use of the theory of integral equation, we introduce the following notations:
\begin{equation*}
\genfrac{}{}{0pt}{}{\Omega_1}{\Omega_2} = \int_{0}^{1}  I_1(\alpha t) z \re^{-z/2} \genfrac{}{}{0pt}{}{M(\bar{a},2,z)}{U(\bar{a},2,z)} \rd t,
\end{equation*}
and
\begin{equation*}
\genfrac{}{}{0pt}{}{\Omega_3}{\Omega_4} = \int_{0}^{1}  K_1(\alpha t) z \re^{-z/2} \genfrac{}{}{0pt}{}{M(\bar{a},2,z)}{U(\bar{a},2,z)} \rd t,
\end{equation*}
where $\bar{a}=a(c=0)$. In addition, denote the Wronskian of any three linearly independent solutions, $f_1,f_2$ and $f_3$ by $\Lambda[f_1,f_2,f_3]$. Then we have the shorthand notations:
\begin{align*}
\Lambda_1(t){\equiv}\Lambda &[I_1(\alt),K_1(\alt),\phione(t)],&&\Lambda_2(t){\equiv}\Lambda[I_1(\alt),K_1(\alt),\phitwo(t)], \\
\Lambda_3(t){\equiv}\Lambda &[I_1(\alt),\phione(t),\phitwo(t)],&&\Lambda_4(t){\equiv}\Lambda[K_1(\alt),\phione(t),\phitwo(t)]. 
\end{align*}

Thus stationary Green's function, $G_0(r,t)=g_0(r,t,c=0)$, is found to be 
\begin{equation} \label{eq:Greensn0}
G_0(r,t)=d_0 
\begin{cases}
 [ \Omega_0 \Lambda_1(t) + \Omega_3 \Lambda_3(t) + \Omega_1 \Lambda_4(t)] I_1(\alr) \; + \\ \;\;\;\;  [\Omega_2 \Lambda_1(t) + \Omega_1 \Lambda_2(t) + \Lambda_3(t) ]\phione(r), \;\;\;\;\;\; 0 \leq r \leq t, \\
 & \\
 [\Omega_0 \Lambda_1(r) +\Omega_3 \Lambda_3(r) ]I_1(\alt) + \Omega_1 \Lambda_3(r) K_1(\alt) \; + \\
 \;\;\;\; [\Omega_2 \Lambda_1(r)-\Lambda_3(r)]\phione(t) + \Omega_1 \Lambda_1(r) \phitwo(t), \;\;\;\;\;\; t \leq r \leq 1, \\
\end{cases}
\end{equation}
where $d_0{=}-1/(\Omega_1 W_0)$, and $\Omega_0{=}\Omega_1 \Omega_4 {-} \Omega_2 \Omega_3$.
Equation (\ref{psi0}) may be transformed into a Fredholm integral equation of the second kind
\begin{equation} \label{psi0-fred}
	\psir(r) + \lambda \int_0^1   K_0(r,t)\psir(t) \rd t =0, 
\end{equation}
where the kernel is given by
\begin{equation*}
	K_0(r,t)=\Big( {\partial^2 (G_0 t^3)}/{\partial^2 t} - {\partial (G_0 t^2)}/{\partial t} - G_0 (\alpha^2 t^3 + t) \Big)/4.
\end{equation*}
We have $G_0(r,t)=G^{*}_0(t,r)$, in view of the reciprocity theorem of Green's functions. Also $K_0(0,t)=K_0(1,t)=0$. The properties of $G_0$ guarantee that $K_0(r,t)$ is continuous for $0 < r,t \leq 1$. For finite $\alpha$ and $R$, $\abs{G_0}$ is clearly bounded from the known asymptotic properties of the confluence hypergeometric functions. 

Hence $\abs{K_0} < N_1$, say. Moreover, $\partial K_0 /{\partial t}$ is continuous for all $r \neq t$ and the only discontinuity at $t=r$ is simply
\begin{equation*}
  { \partial (r^3 K_0) }/{\partial t} \bigr\rvert_{t=r+0}-{ \partial (r^3 K_0) }/{\partial t} \bigr\rvert_{t=r-0}=-1.
\end{equation*}
Accordingly $K_{0}$ satisfies the Lipschitz condition of
\begin{equation*}
  \abs{ K_{0}(r,t_{1})-K_{0}(r,t_{2}) }\;{\leq}\;M_1\; \abs{ t_{1}-t_{2} },
\end{equation*}
where $M_1$ is a constant.
It follows, from standard theory of Fredholm integral equation, that $\Delta_M$ is an entire function of $c$ for finite $\alpha$ and $R$. 
For given $\alpha=\alpha_1$ and $R=R_1$, let the integral, 
\begin{equation*}
\int_0^1 \int_0^1 K_0(r,t,\alpha,R)K_0(t,r,\alpha,R)\rd t \rd r,
\end{equation*}
vanish. Since $K_0$ is a continuous function of these parameters, there must exist a pair of $\alpha \neq \alpha_1$ and $R \neq R_1$, which render the above integral non-zero. Otherwise, if the integral vanishes for all the parameters, $K_0$ must be a constant. Hence $\Delta_M$ has at least one non-zero eigenvalue; the spectrum of $\Delta_M$ is never empty. By an analogous procedure for $\Delta_T$, we may show that the entire function $\Delta_M$ has an order of at most $2/3$. As $K_0(r,t)$ cannot be degenerated, {\it the spectrum of $\Delta_M$ must consist of infinitely many eigen-modes for given $\alpha$ and $R$. Every solution of $\Delta_M$ is only possible for $c_i \leq -\alpha/R$}, as shown in the previous section.
Finally, we notice that the elements of $\Delta_M$ are necessarily distinguish from those of $\Delta_T$.
\subsection*{Asymptotic properties}
By a change of independent variable $y=r^2$, equation (\ref{whit0-phi}) is transformed into
\begin{equation*}
	\phi''(y)+ Q^2 \Big( 1+(\bar{c} -1 )/y \Big) \phi(y)=0, 
\end{equation*}
where $Q=\sqrt{\ri \alpha R}/2$, and $\bar{c}=c+\ri \alpha/R$. According to the Liouville-Green or the WKBJ procedures (see, for example, Bender \& Orszag 1978; Olver 1997), we have, as $\alpha R \rightarrow \infty$,
\begin{equation*}
	Q\int_0^1  \Big( 1 + \frac{\bar{c}-1}{y} \Big) ^{1/2} \rd y = Q \Big(\: \sqrt{ \bar{c}} +(\bar{c}-1)\log \Big(\: \frac{1+\sqrt{ \bar{c} }}{\sqrt{ \bar{c} -1}} \:\Big) \:\Big)
\end{equation*}
by direct evaluation.
If $\abs{\sqrt{\bar{c}}} > 1$, the terms containing $\bar{c}$ can be simplified as
\begin{equation*}
	2\sqrt{\bar{c}}-2/(3\sqrt{ \bar{c}}) + \mbox{terms of} \;\;1/{\bar{c}}^{3/2}, 1/\bar{c}^{2}, \; \cdots .
\end{equation*}
Then the set of the spectrum for the mean modes is given by
\begin{equation} \label{n0-meanmodes}
	c={2}/{3}-\ri \big( m^2 \pi^2 + \alpha^2 \big)/(\alpha R),\;\;\;\;\;\;(m=1,2,\cdots). 
\end{equation}
On the other hand, if we are interested in what happens close to the pipe wall, we may ignore any contributions from the pipe centre in the integral. The terms in $\bar{c}$ reduce to, for $\abs{\bar{c}} \ll 1$
\begin{equation*}
	\sqrt{ \bar{c}} +(\bar{c}-1)\log(1+\sqrt{\bar{c}}) \; \approx \;{\bar{c}}^{3/2}.
\end{equation*}
Hence the following formula furnishes a leading order approximation for the wall modes
\begin{equation} \label{n0-wallmodes}
	c = {(2 \pi l)}^{2/3}  \re^{-\ri \pi/6} /(\alpha R)^{1/3} - \ri \alpha^2/(\alpha R),\;\;\;\;\;\;(l=1,2, \cdots ).
\end{equation}
Since equation (\ref{whit0}) possesses an irregular singularity of rank $1$ at $\abs{z} \rightarrow {\infty}$, formal solutions can be constructed for Kummer's function, see (\ref{eq:m-asymp}). For large $\abs{a}$, the wall modes can be calculated with the aid of (\ref{eq:m-asymp}) as long as sufficient terms in the sums are included. In particular, the eigenvalue relation has the first order approximation of
\begin{equation} \label{n0-centre}
	\Delta_T \sim \re^{\ri {\pi}a} {z_0^{1-a}} {\re^{-z_0/2}}/ \big(\: (1-a)\Gamma(1-a)\:\big) + z_0^{a-1} \re^{z_0/2}/\Gamma(a).
\end{equation}
The first term is dominant and the second recessive {\it provided that} $\Real (a)$ is small. As $\abs{1/\Gamma(1-a)} \rightarrow 0$, $\Delta_T$ is exponentially small. Hence we have
\begin{equation} \label{n0-centremodes}	
c=1-{4k} \: \re^{\ri{\pi}/4}/({\alpha R})^{1/2}-\ri {\alpha}^{2}/({\alpha R}),\;\;\;(k=1,2, \cdots ).
\end{equation}
As $\abs{c}$ increases, so does $\Real (a)$, then the magnitude of $z_0^{1-a}$  decreases until the first term in (\ref{n0-centre}) becomes numerically small; the reverse holds in the second term. When $z_0^{1-a}\sim O(\re^{z_0/2})$, the nearest integer of 
\begin{equation} \label{maxk} 
	\big| z_0/(2\log z_0) \big|
\end{equation}
gives a useful indication of the maximum allowable $k$. For $\alpha=1, \alpha R=10^4,10^6$ and $10^8$, there are $10,69$ and $526$ (torsional) centre modes respectively. In solution (\ref{phir0-volt-sol}), ${\psir}_1 \sim \psione$, and ${\psir}_3 \sim f_3$ for large $\alpha R$, then the dispersion law (\ref{mm-modes}) is approximately satisfied by $\psione(1)=0$ and $\psione'(1)=0$. Within a small error bound, we find that the centre modes of $\Delta_M$ coincide with those of $\Delta_{T}$, see also Pekeris (1948). By repeating the LG procedures with the modification of the wall boundary condition, we assert that the mean modes of $\Delta_{M}$ are given by
\begin{equation*}
	{2}/{3}-\ri \big( (m-1/2)^2 \pi^2 + \alpha^2 \big) / (\alpha R),\;\;\;\;\;\;(m=1,2, \cdots ).
\end{equation*}
The following expression which is valid for $\alpha \ll R$ furnishes an asymptotic approximation for the wall modes of $\Delta_M$ 
\begin{equation*}
	-2^{2/3}q_{\pm s}\:\re^{-\ri \pi/6} /(\alpha R)^{1/3} , \;\;\;\;\;\;(s=1,2,\cdots ),
\end{equation*}
where $q_{\pm s}$ denote the (complex) zeros of the integral of Airy's function. The formula was first given by Corcos \& Sellars (1959) and further justified by Gill (1965).
\section{Analysis of the general case $n \geq 1$}
In view of the variation of parameters, the general solution of equation (\ref{eq:pressure}) may be expressed as
\begin{equation} \label{press-sol}
	p(r)=A_{0} I_{n}(r) + B_{0} K_{n}(r) -4{\alpha}^{2} \int_{0}^{r}  Y(r,t)\psi_{r}(t) \rd t,
\end{equation}
where $A_{0}$, $B_{0}$ are constants, and 
\begin{equation*}
Y(r,t)=\Big( I_{n}(\alr)K_{n}(\alt) - I_{n}(\alt)K_{n}(\alr) \Big)t^{2}.
\end{equation*}
The functions, $I_{n}({\alpha}r)$ and $K_{n}({\alpha}r)$, are the modified Bessel functions of the first and second kinds. By the recurrence and derivative properties of the Bessel functions, we derive the following driving functions from the pressure solution (\ref{press-sol}):
\begin{equation} \label{ppm}
  p\:'(r){\mp}\frac{n}{r}p(r)=A_{0} I_{n {\pm} 1}(r)+ B_{0}K_{n{\pm}1}(r) - 4{\alpha}^3 \! \int_{0}^{r}   \! Y_{\pm}(r,t)\psi_{r}(t) \rd t = p_{\mp}(r),
\end{equation}
where again $A_{0}$ and $B_{0}$ are constants. The kernel functions, $Y_{\pm}(r,t)$, have the form of
\begin{equation*}
  Y_{\pm}(r,t)=\Big( I_{n{\pm}1}(\alr)K_{n}(\alt)+I_{n}(\alt)K_{n{\pm}1}(\alr) \Big) t^{2}. 
\end{equation*}
The unknown functions of (\ref{u-comps}), $\psir(r)$ and $\psit(r)$, can be decoupled by introducing two auxiliary functions
\begin{equation*}
  \psi_{\pm}(r)=\psi_{r}(r)\;{\pm}\;\psi_{\theta}(r).
\end{equation*}
Adding and subtracting the second and the third equations of (\ref{u-comps}) yield the following {\em coupled} differential system
\begin{equation} \label{psi-pm}
  \psi''_{\pm} + \psi'_{\pm}/r - \Big( \alpha^2 + (n \pm 1)^2/r^2 + \ri \alpha R (1-r^2-c) \Big) \psi_{\pm}=-\ri Rp_{\mp}(r). 
\end{equation}
The appropriate boundary conditions for $\psi_{\pm}(r)$ follow directly from the boundary conditions for $\psir(r)$ and $\psit(r)$.
Denoting $\phi_{\pm}(r)=r\psi_{\pm}(r)$, the coupled system in (\ref{psi-pm}) may be simplified as 
\begin{equation} \label{phi-pm}
  \phi''_{\pm} - \phi'_{\pm}/r - \Big( \alpha^2+ \big((n \pm 1)^2-1 \big)/r^2 + \ri \alpha R (1-r^2-c) \Big) 
\phi_{\pm}=-\ri Rr p_{\mp},
\end{equation}
which can be further transformed into a coupled ordinary differential equations with independent variable $z$:
\begin{equation*} 
\phi''_{\pm}(z) + \Big( -1/4+\kappa/z + \big( 1-(n \pm 1)^2 \big)/(4z^2) \Big) \phi_{\pm}(z)=p_{\mp}/(4 \alr). 
\end{equation*}
These are Whittaker's equation. The auxiliary functions are readily expressed as 
\begin{equation} \label{psi-pm-sol}
\psi_{\pm}(r) = A_{\pm}M_{\pm}(r)+B_{\pm}W_{\pm}(r)-\frac{1}{8{\alpha} z_0 }\frac{\Gamma(a_{\pm})}{\Gamma(b_{\pm})}  \int_{0}^{r}   G_{\pm}(r,t)p_{\mp}(t) \rd t,
\end{equation}
where $A_{\pm}$ and $B_{\pm}$ are arbitrary constants. 
The shorthand functions, $M_{\pm}$ and $W_{\pm}$, are related to the standard confluent hypergeometric functions by 
\begin{equation*}
\begin{split}
  M_{\pm}(r) & =z^{b_{\pm}/2} \re ^{-z/2}M(a_{\pm},b_{\pm},z)/r,\\
  W_{\pm}(r) & =z^{b_{\pm}/2} \re ^{-z/2}U(a_{\pm},b_{\pm},z)/r,
\end{split}
\end{equation*}
where
\begin{equation*}
  b_{+}=n+2,\;\;\; b_{-} = n,\;\;\; a_{\pm} = b_{\pm}/2 - \kappa.
\end{equation*}
The kernel functions, $G_{\pm}(r,t)$, are given by
\begin{equation*} 
  G_{\pm}(r,t)=\Big( M_{\pm}(r)W_{\pm}(t) - M_{\pm}(t)W_{\pm}(r)\Big)/t. 
\end{equation*}

Substituting (\ref{ppm}) for $p_{\mp}$ of (\ref{psi-pm-sol}), the auxiliary functions may be converted into a Volterra integral equation of the second kind for the radial velocity component, namely,
\begin{equation} \label{psi-volt}
  \psi_{r}(r)-z_0^{-1}\int_{0}^{r} \! K(r,t)\psi_{r}(t) \rd t = F(r), 
\end{equation}
where the kernel has the integral expressions of
\begin{equation*} 
\begin{split}
K(r,t) & = C_{+}   \int_{t}^{r}   G_{+}(r,s)Y_{+}(s,t) \rd s + C_{-}   \int_{t}^{r}   G_{-}(r,s)Y_{-}(s,t) \rd s \\
 & = C_{+} K_{+} + C_{-} K_{-}, \\
 \end{split}
\end{equation*}
where $C_{\pm} = {\alpha^2}{\Gamma(a_{\pm})}/\big( 4{{\Gamma(b_{\pm})} }\big)$.
Introducing the following shorthands,  
\begin{equation} \label{ed-volt}
\begin{split}
D_{\pm}(r) & = C_{+}   \int_{0}^{r} G_{+}(r,t)I_{n+1}(\alt) \rd t\;{\pm}\; C_{-}   \int_{0}^{r} G_{-}(r,t)I_{n-1}(\alt) \rd t, \\
E_{\pm}(r) & = C_{+}   \int_{0}^{r} G_{+}(r,t)K_{n+1}(\alt) \rd t\;{\pm}\; C_{-}   \int_{0}^{r} G_{-}(r,t)K_{n-1}(\alt) \rd t, 
\end{split}
\end{equation}
the right-hand function $F$ is expressed in terms of the shorthand functions as
\begin{equation*} 
  F = A_{+}M_{+} + A_{-}M_{-} + B_{+}W_{+} + B_{-}W_{-} + A_0 D_{+} + B_0 E_{+}.
\end{equation*}
By the method of successive approximations, the solution $\psi_{r}$ is given by
\begin{equation} \label{psi-soln}
	\psi_{r}(r)=F(r)+\int_{0}^{r} \! H(r,t)F(t)  \rd t, 
\end{equation}
where the resolvent kernel $H(r,t)$ has the expansion in terms of the iterated $K_j$ (cf. (\ref{iter})). From appendix A, the asymptotic properties of Whittaker's functions ensure that $\abs{K(r,t)} < N$ for some constant $N$. Note that $\int_0^r H(r,t) D_+(t) \rd t$
is bounded. The solutions of $\psir(r)$ (as well as $\psir'(r)$) are everywhere regular because the product of $H(r,t)$ and any one of the bounded parts of $F(t)$ is regular over $0 \leq t,r \leq 1$. 

To simplify the subsequent presentation, we introduce the following notations for any continuous function $f(r)$
\begin{equation*} 
\begin{split}
	\psi[f(r)]&=f(r)+ \int_{0}^{r} H(r,q) f(q) \rd q, \\
	\psi'[f(t)]&=f'(r) \bigr\rvert_{r=t} +\int_{0}^{t} \frac{\partial H(r,q)}{\partial r}\bigr\rvert_{r=t}\:f(q) \rd q.\\
	\end{split}
\end{equation*}
Then function $\psi'_r$ can be computed from
\begin{equation*} 
	\psi'_{r}(r)=\psi'[\psir(r)]. 
\end{equation*}

Once the solutions for $p_{\mp}$ have been obtained in terms of the solutions of $\psi_{r}$, the auxiliary functions $\psi_{\pm}$ can be written as
\begin{equation*}
\begin{split}
	\psi_{\pm}(r)/2 & =A_{\pm}M_{\pm}(r) +B_{\pm}W_{\pm}(r) +A_{0} C_{\pm} \int_{0}^{r}   G_{\pm}I_{n{\pm}1}(\alt) \rd t \\
	& \quad  + B_{0} C_{\pm}   \int_{0}^{r}   G_{\pm}K_{n{\pm}1}(\alt) \rd t +\int_{0}^{r}   K_{\pm}(r,t) \psi[F(t)] \rd t.
	\end{split}
\end{equation*}
Specifically, the solution for the circumferential disturbance component is found to be
\begin{equation*} 
\begin{split}
	\psi_{\theta}(r) & =A_{+}M_{+}(r)-A_{-}M_{-}(r)+B_{+}W_{+}(r)-B_{-}W_{-}(r) + \\
	& \quad A_{0} D_{-}(r) + B_{0} E_{-}(r) +\int_{0}^{r} J(r,t)\psi[F(t)] \rd t,
	\end{split}
\end{equation*}
where 
\begin{equation*}
 J(r,t)=C_+ K_{+}(r,t)- C_- K_{-}(r,t). 
\end{equation*}
From the equation of continuity, $\psix$ is also known and can be expressed in terms of $\psir$, $\psir'$ and $\psit$. We shall refrain writing down the full complicated expression. It can be seen that the linearized equations of motion is equivalent to a {\it sixth} order system. For bounded velocities and pressure in $0 \leq r \leq 1$ we require that $B_{\pm}=0$ and $B_{0}=0$. Applying the boundary conditions at the pipe wall for $\psi_{r}$, $\psi'_{r}$ and $\psi_{\theta}$, we obtain $3$ homogeneous linear equations for the constants $A_{\pm}$ and $A_{0}$.
The secular determinant defines the dispersion relation as follows.
\begin{equation} \label{eigen}
\Delta_n(\alpha,R,n,c)=\left| \begin{array}{ccc}
     a_{11}       &  a_{12}  &  a_{13}   \\
     a_{21}&  a_{22}  &  a_{23}   \\
     a_{31} & a_{32} & a_{33}  
                      \end{array}    \right|=0,               
\end{equation}
where $a$'s are the entire functions of the parameters $\alpha, R, n$, and they are given by
\begin{equation} \label{eigen-as}
\begin{split}
a_{11} & = \psi[M_{+}(1)],\;\;\; a_{12} = \psi[M_{-}(1)], \;\;\; a_{13} =  \psi[D_{+}(1)], \\
& \\
a_{21} & = \psi'[M_{+}(1)],\;\;\; a_{22} = \psi'[M_{-}(1)], \;\;\; a_{23}=  \psi'[D_{+}(1)], \\
& \\
 a_{31}  & =  M_{+}(1) + \int_{0}^{1}  \! J(1,t) \psi[ M_{+}(t)] \rd t, \\ 
 a_{32}  & = -M_{-}(1) + \int_{0}^{1}  \! J(1,t) \psi[ M_{-}(t)] \rd t, \\
 a_{33}  & =  D_{-}(1) + \int_{0}^{1}  \! J(1,t) \psi[ D_{+}(t)] \rd t. \\
\end{split}
\end{equation}

The stability defined in (\ref{eigen}) can be established in several steps. First, because $\psir(1)$, $\psit(1)$ and $\psix(1)$ (or $\psir'(1)$) are the linearly independent boundary conditions imposed at the pipe wall, any row (or column) cannot be proportional (or equal) to any other row (or column). Nor can any sum (or difference) of two or three rows (columns) be proportional (or equal) any other row (column). It is also impossible for two or three minors of a row (column) to vanish, as it implies that two of the three boundary conditions are linearly dependent. For instance, the minors of $a_{11}$ and $a_{12}$ vanish or $M_{11}=0$ and $M_{12}=0$. Then $a_{22}/a_{23}=a_{32}/a_{33}=k$, say. Also $a_{21}/a_{23}=a_{31}/a_{33}=k$ so that the second row would be proportional to the third. Moreover, if one of the minor vanishes, take $M_{11}=0$, then $\Delta_n=M_{22}(a_{13}k-a_{12})$. For $\Delta_n$ to vanish, either $M_{22}=0$ or $a_{12}=k a_{13}$. Both the conditions indicate $\psir'(1)$ and $\psit(1)$ are linearly dependent. Similar discussions hold for other minors of any rows or columns. In brief, no row or column of any minors can be equal or in proportion. Second, if $c_i > -\alpha/R$, none of $M_{\pm}(1)$ and $M'_{\pm}(1)$ can vanish in the light of the analysis given in the Appendix A. Now assume $a_{11}=0$, the integrand in $a_{11}$, for all $t$ in $0$ to $1$,
\begin{equation*}
 M_{+}(t)H(1,t)=-M'_{+}(t)
\end{equation*}
by the fundamental theorems of calculus since $M_+(0)=0$. This identity does not hold at $t=1$ as $H(1,1) \equiv 0$ and we have a contradiction. So $a_{11} \neq 0$. Similarly $a_{12} \neq 0$ in general. The only exception is that $M_-(0)=z_0$ at $n=1$. In this case, $a_{12}$ can only vanish if the integrand in $a_{12}$ equals $-M'_-(t)-z_0$. Because $M'_-(1) \neq -z_0$ for $\alpha R > 0$ it follows that $a_{12} \neq 0$. Since $J(1,1) \equiv 0$, we see that both $a_{31}$ and $a_{32}$ cannot vanish by analogous arguments. Moreover, the derivative properties of $M_{\pm}$ show that if $M''_{+}(1)$ (or $M''_{-}(1)$) vanishes then $M''_{-}(1)$ (or $M''_{+}(1)$) cannot vanish. As $\partial{H(r,1)}/{\partial r}(r=1){\equiv}0$, we see that either $a_{21}$ or $a_{22}$ may vanish. Briefly, only one of the six elements in the first two columns may vanish. Next, equation (\ref{ed-volt}) shows that $D_+(0)=D_-(0)=0$. A straightforward analysis demonstrates that $D'_+(0)=0$. For $a_{13},a_{23}$ and $a_{33}$ to vanish, we must require $D'_+(1),D''_+(1)$ and $D'_-(1)$ to vanish respectively. From the derivative properties of Whittaker's functions, we see that at most two of these three requirements can be satisfied. It follows that at most two of the elements in the last column may be zero. In summary, no three elements in any row or any column of (\ref{eigen}) can all vanish. Last, suppose that $a_{23}$ and $a_{33}$ both vanish, and that the elements of the minor of $a_{13}$ are non-zero, then (\ref{eigen}) can vanish only if the minor vanishes. This is impossible as the vanishing minor would imply the two boundary conditions, as defined by $\psir'(1)$ and $\psit(1)$, are linearly dependent. Obviously, similar arguments apply to the other cases in which any two elements in the last column may vanish. In conclusion, {\it the determinant (\ref{eigen}) admits solutions only when $c_i \leq -\alpha/R$}. By means of the general Fredholm theory, we have just shown the existence of the solutions of the eigenvalue relation. Therefore all hydrodynamically admissible modes of disturbance are stable for finite $\alpha$, $R$ and any integer $n \geq 1$.
\subsection*{Alternative derivation of $\Delta_n$}
To investigate whether (\ref{eigen}) admits any solutions, we eliminate $p$ between the first three equations in (\ref{u-comps}). Thus we arrive at three equations as follows: 
\begin{equation} \label{odes}
\begin{split}
n L_n \psix + 2 \ri R r n \psir - \alpha r D_n \psit + 2n \alpha \psir/r &= \lambda (n \psix - \alpha r \psit), \\
D(L_n \psix ) + \alpha D_n \psir + 2 \ri R D(r \psir) - 2 n \alpha \psit /r^2 &= \lambda (\psix' + \alpha \psir), \\
D(rD_n \psit) +  n D_n \psir - 2n \big( D(\psir/r) + n \psit/r^2 \big) &= \lambda (r \psit' + \psit + n \psir ). 
\end{split}
\end{equation}
where 
\begin{equation*}
 L_n=D^2+1/rD-\alpha^2 -n^2/r^2 - \ri \alpha R (1-r^2), \;\;\; D_n=L_n-1/r^2.
\end{equation*}
The first equation is equivalent to an equation in $\psir'''$ from the continuity.
Green's function for (\ref{odes}), that is, for the operators in (\ref{u-comps}) together with the boundary conditions consists of nine components. Denote them by $g_k(r,t,c)$, for $k=1,2,3$. They have the form of
\begin{equation} \label{greens-gn}
g_k(r,t,c)=
\begin{cases}
\:\overset{6}{\underset{j=1}\sum} A_{j}^{k}(t) {\psi}_j(r), & \; 0 \leq r \leq t, \\
\:\overset{6}{\underset{j=1}\sum} B_{j}^{k}(t) {\psi}_j(r), & \; t \leq r \leq 1,\\
\end{cases}
\end{equation}
where $\psi_j$ stands for ${\psir}_j$, ${\psix}_j$ or ${\psit}_j$, the subscripts refer to the six linearly independent solutions, and $A$'s and $B$'s are the unknown functions to be determined. From the properties of the solutions at the pipe centre, we require that $A_2=A_4=A_6=0$ for bounded solutions. For the nine remaining unknowns, three conditions come from the wall boundary condition for $g_k$, in other words, for $\psir$, $\psix(\psir')$ and $\psit$ at $r=1$. At $r=t$, we have two continuity conditions for $g_k$ and $g'_k$. The nature of the equation of continuity shows that these two conditions effectively define five algebraic equations. Finally, we have a jump condition in $g_k''$. 
In particular, the three components for $k=1$ are all zero. Other components are found and given in appendix B. The system of the governing equations (\ref{u-comps}) is therefore equivalent to the Fredholm integral equation of the second kind, from (\ref{eq:sys-intgleqns}),
\begin{equation*}
\uppsi(r)-{\lambda}\int_{0}^{1}   N(r,t)\:{\uppsi}(t)\: \rd t=0,
\end{equation*}
where $\uppsi$ stands for the column vector containing $(\psir,\psix,\psit )$ and the kernel $N$ is the $3{\times}3$ matrix whose elements $N^{*}_{ij}(t,r)=N_{ij}(r,t)$. As we have discussed for the case of $n=0$, $\abs{N}$ possesses an upper bound, say $N_n$. The definitions of the components of Green's function guarantee that $N(r,t)$ is everywhere continuous in $r$ and $t$. The jump conditions at $t=r$ can only result in $N$ with {\em finite discontinuities} and hence $N(r,t)$ can be considered to satisfy the Lipschitz condition of unity. By analogy with the standard Fredholm theory, it is seen that (\ref{eigen}) is an entire function of the parameters.  For given finite $\alpha=\alpha_1$ and $R=R_1$ at fixed $n$, if the integral, 
\begin{equation*}
\int_0^1 \int_0^1 N(r,t,\alpha,R,n)N(t,r,\alpha,R,n)\rd t \rd r,
\end{equation*}
vanishes, it implies that there exists a pair of $\alpha$ and $R$ which ensures that the integral is non-zero. In other words, if the integral vanishes for all values of $\alpha$, $R$ and $c$, the continuous function $K$ must be a constant. It follows that (\ref{eigen}) does admit solutions and that there exists at least one non-zero eigenvalue. The spectrum $\Delta_{n}$ is non-empty. It is evident that $N(r,t)$ cannot be degenerated into any form of $\sum_{j=1}^{k} S_j(r) T_j(t)$ for finite $k$ and for some functions $S$ and $T$. The spectrum of (\ref{eigen}) consists of infinitely many eigenvalues because the order of this entire function is at most $2/3$. The present analysis dose not seem to lend itself for definite assessment of the (finite) multiplicity and degeneracy of the eigenvalues. 
\subsection*{Asymptotic approximation}
When $\abs{z_0}$ becomes large, it may be shown that the contributions from the resolvent kernel $H$ become insignificant and therefore may be neglected. Then the elements of the first row of (\ref{eigen}) vanish and the eigenvalues are defined by
\begin{equation} \label{n-tmodes}
\Delta^{T}_n(\alpha,R,c_{\pm})=z_0^{b_{\pm}/2} \re^{- z_0 /2} M(a_{\pm},b_{\pm},z_{0})=0
\end{equation}
and
\begin{equation} \label{n-mmodes}
\Delta^{M}_n(\alpha,R,c_{\pm})=z_0^{b_{\pm}/2}\int_{0}^{1} I_{n{\pm}1}(\alpha r)r^{b_{\pm}}\re^{-z_0r^2/2}M(a_{\pm},b_{\pm},z_0r^2) \rd r=0.
\end{equation}
In each case, we have used the $\pm$ subscripts to distinguish the two possible values of $c$. In fact, these two values are very close to each other. Either of them or the average of them may be used as an approximation. If $\alpha R \gg n$, the relation (\ref{n0-meanmodes}) for the mean modes still holds because the last terms on the left hand side of Whittaker's equation (\ref{whit}) may be neglected.
Furthermore, the asymptotic approximation to the wall modes of (\ref{n-tmodes}) can be obtained from the left-hand side of (\ref{phi-pm}), $\phi_{\pm} \rightarrow \phi_n$, namely,  
\begin{equation*} 
\phi''_n - \phi'_n/r - \big( \alpha^2 + n^2/r^2 + \ri \alpha R (1-r^2-c) \big) 
\phi_n=0.
\end{equation*}
Carrying out a co-ordinate perturbation, $r=1-y$, for small $y$ compared to unity, and neglecting all terms containing $y^2$, the equation is further simplified to
\begin{equation*}
S''-\big( 2\ri \alpha R \eta + (2y-1)/4 \big) S=0,
\end{equation*}
where $S(y)=\phi_n \exp(y/2+y^2/4)$, and 
\begin{equation*}
\eta=y \big(1-\ri n^2/(\alpha R)\big)-\big(c+\ri (\alpha^2+n^2) / (\alpha R)\big)/2. 
\end{equation*}
For large $\alpha R$, the second term in the brackets in the differential equation may be ignored compared to the first one. By neglecting the second linearly independent solution which is expected to be recessive near the wall, we obtain
\begin{equation} \label{n-wall-eigfunc}
S\;{\propto}\;A_i \big( \:(2\ri \alpha R)^{1/3}\eta \:\big),
\end{equation}
where $A_{i}$ is Airy's function. The wall modes are given by, after applying the wall boundary condition, $\eta_s=\eta(y=0)$,
\begin{equation} \label{n-wall-tmodes-Airy}
 - 2^{2/3} r_s \: \re^{-\ri \pi/6} /{(\alpha R)^{1/3}}  - \ri ( \alpha^2 + n^2 )/(\alpha R),\;\;\;(s=1,2, \cdots ),
\end{equation}
where $r_s$ denote the (real) zeros of Airy's function. (The first four zeros are $-2.3381$, $-4.0879$, $-5.5206$ and $-6.7867$ respectively.)
For small $\alpha$ and lower $n$, the modified Bessel functions $I_n(\alpha r)$ in (\ref{n-mmodes}) would have limited effects on the integrals by virtue of the mean value theorem. This suggests that the type of solution (\ref{n-wall-eigfunc}) may be used to approximate Whittaker's function 
\begin{equation*} \label{n-mmodes-approx}
\Delta^{M}_n(\alpha,R,c_{\pm}) \propto I_n(\alpha) \:B \int_{\eta_1}^{\eta_s}   A_i\{(2\ri \alpha R)^{1/3} \eta\} \rd \eta,
\end{equation*}
where the lower integration limit $\eta_1$ corresponds to $y \rightarrow 1$. The factor $B$ depends on the slowly varying function $\exp(y/2+y^2/4)$. Provided that Airy's function is exponentially small at $\eta_1$, the wall modes may be approximated by
\begin{equation} \label{n-wall-mmodes-Airy}
	- 2^{2/3} q_{\pm s} \: \re^{-\ri \pi/6} /{(\alpha R)^{1/3}}  - \ri ( \alpha^2 + n^2 )/(\alpha R),\;\;\;(s=1,2, \cdots ).
\end{equation}
On the other hand, as $\alpha$ and $n$ become large, $I_n(\alpha r)$ tends to $e^{\alpha r}/\sqrt{2 \pi \alpha r}$ and $(\alpha r)^{n}/(2^n n!)$ respectively. It is plausible that the principle contribution to the integrals in (\ref{n-mmodes}) must come from a small interval close to the upper limit. We have for some $\zeta \approx 1$,
\begin{equation} \label{n-mmodes-by-tmodes}
\Delta^{M}_n(\alpha,R,c_{\pm}) \approx  z_0^{b_{\pm}/2} \zeta^{b_{\pm}} \re^{- z_0 \zeta^2 /2 } M(a_{\pm},b_{\pm},z_0{\zeta}^2)=0.
\end{equation}
Hence we expect a somewhat similar distribution of the wall modes as that given by (\ref{n-wall-tmodes-Airy}). 
If $\Real (a_{\pm})$ are small (see appendix A), the relations in (\ref{n-tmodes}) have asymptotic expansions of 
\begin{equation}
\re^{\ri {\pi}a_{\pm}} {z_0^{ b_{\pm}/2 - a_{\pm}}} \re^{-z_0 / 2} \Gamma(b_{\pm})/ \Gamma(b_{\pm}-a_{\pm}).
\end{equation}
For fixed $\alpha R$, if $n$ is so large that $\Gamma(b_{\pm})$ are sufficiently large, then these expansions define no eigen-modes; there exist no centre modes at large $n$. For small and moderate $n$, they are exponentially large unless 
\begin{equation} \label{centre-cond}
a_{\pm}-b_{\pm}=k,\;\;\;k=0,1,2  \cdots .
\end{equation}
The maximum allowable $k$ is not greater than that of (\ref{maxk}). Then a pair of the centre modes is given by, as $\alpha R \rightarrow \infty$, 
\begin{equation} \label{n-centremodes}
\begin{split}
c_{+}&=1-{2(2k+n)}\:\re^{{\pi}\ri/4}/\sqrt{\alpha R} -\ri {{\alpha}^{2}}/(\alpha R),\\
c_{-}&=1-{2(2k+n+2)}\:\re^{{\pi}i/4}/\sqrt{\alpha R} -\ri {{\alpha}^{2}}/(\alpha R).
\end{split}
\end{equation}
As $a_{\pm}$ or $\kappa_{\pm}$ are the only complex quantities, linear combinations of the two relations in (\ref{centre-cond}) require that we have either $\Imag \kappa_+=-\Imag \kappa_-$ or $\Imag \kappa_+=\Imag \kappa_-$. The former is rejected as it implies $c_i \geq 0$. Subtraction between the two relations gives us $
\Real \kappa_+ - \Real \kappa_- = \pm 1$, which give rise to, corresponding to (\ref{n-centremodes}), 
\begin{equation} \label{n-centremodes-ave}
\begin{split}
c_{+}& =1-{{\sqrt{2}}(2k+n+2)}/\sqrt{\alpha R}-\ri \: \Big( {{\sqrt{2}}(2k+n)}/\sqrt{\alpha R}+ {{\alpha}^{2}}/(\alpha R) \Big), \\
c_{-}& =1-{{\sqrt{2}}(2k+n)}/\sqrt{\alpha R}-\ri \: \Big( {{\sqrt{2}}(2k+n+2)}/\sqrt{\alpha R}+ {{\alpha}^{2}}/(\alpha R) \Big).\\
\end{split}
\end{equation} 

To check the various asymptotic formulas, equations in (\ref{u-comps}) have been solved by two complementary numerical schemes. The first one is an improved version of the Chebyshev collocation technique (see, for example, Khorrami {\it et al.}, 1989). In the second method, the solutions are expanded in a power series near the pipe centre and they are then continued by a fourth order Runge-Kutta integrator. Dispersion relation (\ref{eigen}) is established numerically at the pipe wall. Both schemes have been extensively tested and verified against published data throughout the present work. In figures 1 and~2, we present selected computational results. The centre modes are satisfactorily approximated by (\ref{n-centremodes-ave}) for $\alpha \ll R$ and by (\ref{n-centremodes}) for $\alpha \geq R$. The effect of $\alpha$ and $n$, implied in (\ref{n-wall-tmodes-Airy})-(\ref{n-mmodes-by-tmodes}), has been confirmed. For information, the summary below lists computed numerical results for selected wall modes at $\alpha{=}1, \; \alpha R{=}10^4$:
\begin{equation*}
\begin{array}{ccccc}
  \texttt{Dispersion} & \texttt{Method} &\texttt{Eigen-value} \;c\;(n=1) & \texttt{Mode} & \texttt{Figure}\\
  & & & & \\
  (\ref{eigen}) & \texttt{Collocation} & $0.2738-0.0472$\ri & \texttt{Wall} & \texttt{1(b)}\\
  (\ref{n-mmodes}) & \texttt{Asymptotics}\;(c_{+}) & $0.2811-0.0796$\ri & \texttt{Wall} & \texttt{1(b)}\\
  (\ref{n-mmodes}) & \texttt{Asymptotics}\;(c_{-}) & $0.2896-0.0736$\ri & \texttt{Wall} & \texttt{1(b)}\\
    & & & & \\
  (\ref{eigen}) & \texttt{Collocation} & $0.1464-0.0810$\ri & \texttt{Wall} & \texttt{1(b)}\\
  (\ref{n-tmodes}) & \texttt{Asymptotics}\;(c_{+}) & $0.1472-0.0831$\ri & \texttt{Wall} & \texttt{1(b)}\\
  (\ref{n-tmodes}) & \texttt{Asymptotics}\;(c_{-}) & $0.1472-0.0827$\ri & \texttt{Wall} & \texttt{1(b)}\\
  & & & & \\
    \texttt{Dispersion} & \texttt{Method} &\texttt{Eigen-value} \;c\;(n=15) & \texttt{Mode} & \texttt{Figure}\\
  & & & & \\
  (\ref{eigen}) & \texttt{Collocation} & $0.1445-0.1034$\ri & \texttt{Wall} & \texttt{1(f)}\\
  (\ref{n-tmodes}) & \texttt{Asymptotics}\;(c_{+}) & $0.1454-0.1110$\ri & \texttt{Wall} & \texttt{1(f)}\\
  (\ref{n-tmodes}) & \texttt{Asymptotics}\;(c_{-}) & $0.1458-0.1044$\ri & \texttt{Wall} & \texttt{1(f)}\\
\end{array}
\end{equation*}
%
%
\begin{figure}
 \begin{center}
 \includegraphics[width=5.5in,height=6.9in,keepaspectratio]{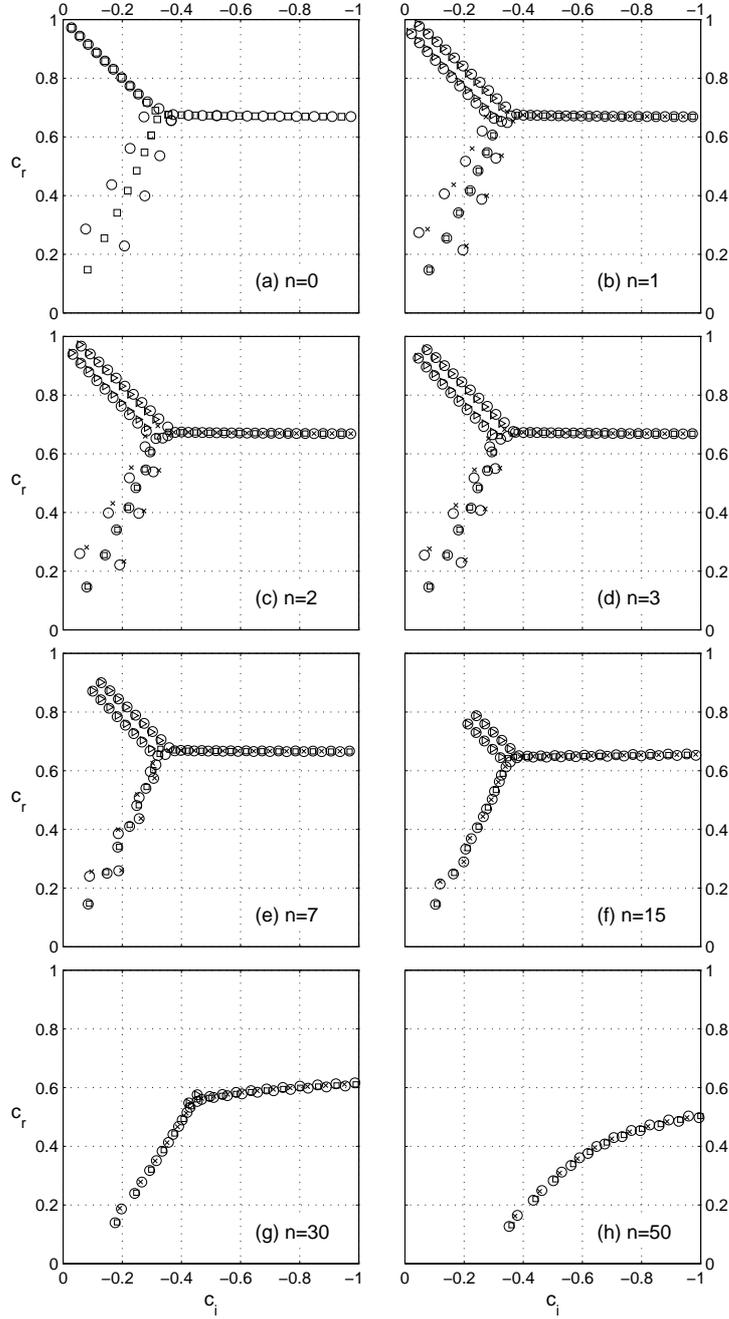}
 \end{center}
\caption{Eigen-mode structure at selected azimuthal periodicity $n$ for ${\alpha}R{=}10^{4}$ and $\alpha{=}1$. (a) Squares ${\Box}$ and circles $\circ$ denote the eigen-modes defined by $\Delta_{T}=0$ and $\Delta_{M}=0$ respectively. The circles in plots (b)-(h) show the results of the Chebyshev collocation method; squares $\Box$ and crosses $\times$ the average of the asymptotic approximations (\ref{n-tmodes}) and (\ref{n-mmodes}), denoted by $c_{\pm}$. Symbols $\rhd$ are the centre modes (\ref{n-centremodes-ave}).}
\end{figure}
\begin{figure}
 \begin{center}
 \includegraphics[width=5.5in,height=6.9in,keepaspectratio]{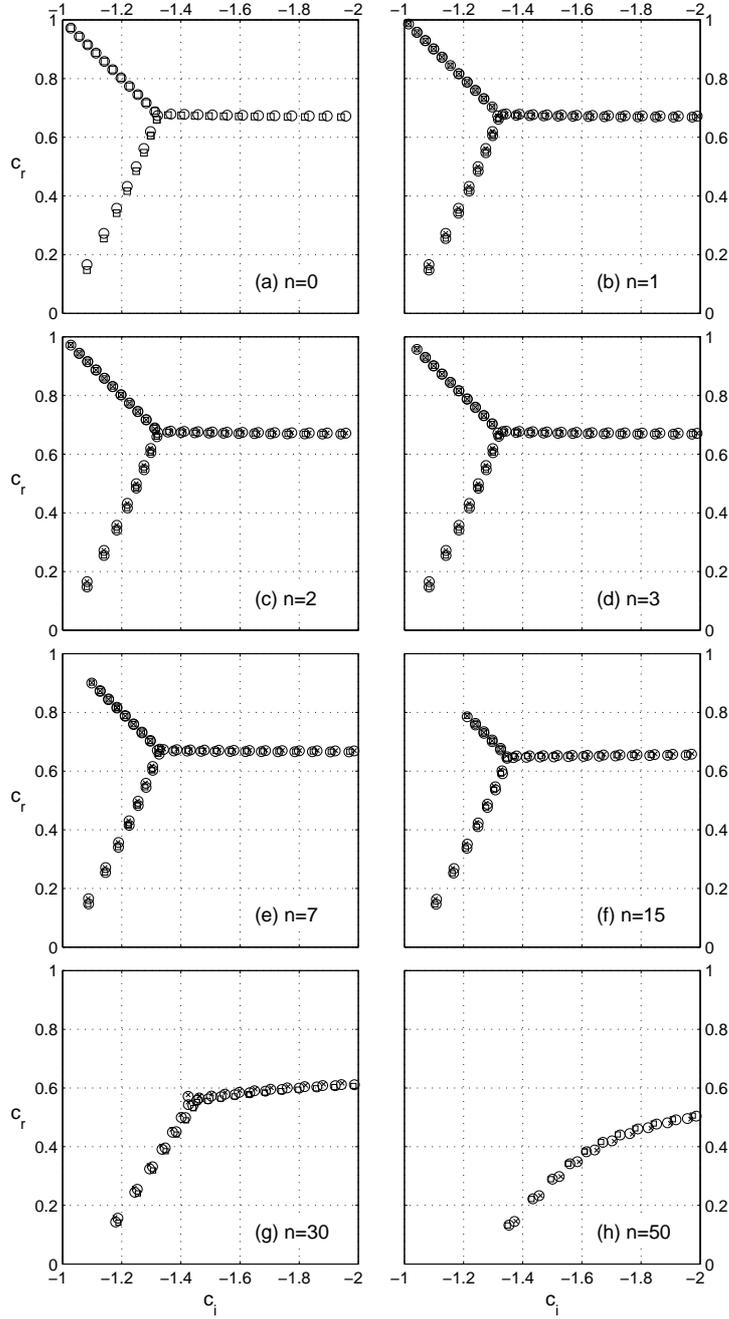}
 \end{center}
\caption{For ${\alpha}R=10^{4}$ and $\alpha=100$. The symbols are identical as in the previous figure. In (b)-(h), the eigen-mode asymptotic approximations are denoted by either $c_+$ or $c_-$. Note the origin of the $c_i$ scale corresponds to $-\alpha/R$.}
\end{figure}
%
%
\section{Limit of $R \rightarrow 0$}
When the viscous force of the fluid motion is far greater than the inertia force, the equations of motion become, for finite $\alpha R$,
\begin{equation} \label{u-comps-r0}
\begin{split}
 \psix'' +\ psix'/r + \big( \beta^2 - n^2 / r^2 \big) \psix & = 0,\\
 \psir'' + \psir'/r + \big( \beta^2 -(n^2+1)/r^2 \big) \psir -2n \psit/r^2 & = 0,\\
 \psit'' + \psit'/r + \big( \beta^2 -(n^2+1)/r^2\big) \psit -2n \psir/r^2 & = 0,\\
 \end{split}
\end{equation}
where $\beta^2=\sigma - \alpha^2$, and the eigenvalue $\sigma=\ri \alpha R c$. Equation (\ref{eq:pressure}) for the pressure remains unchanged. The bounded solutions for the auxiliary equations (\ref{u-comps-r0}) can be expressed in terms of the Bessel functions of the first kind. The solutions for the disturbances have the form of
\begin{equation*}
\begin{split}
  \psir(r) & =  A_{+}J_{n+1}(\beta r)+A_{-}J_{n-1}(\beta r),\\
  \psit(r) & =  A_{+}J_{n+1}(\beta r)-A_{-}J_{n-1}(\beta r),\\
  \psix(r) & =  A_{0}J_{n}(\beta r),\\ 
\end{split}
\end{equation*}
where $A_{\pm}$ and $A_{0}$ are constants. Applying of the wall boundary conditions, we obtain the eigenvalue relation
\begin{equation*}
 \Delta_n(\sigma,\alpha,n)=-2\;J_{n-1}(\beta) \;J_{n}(\beta)\;J_{n+1}(\beta)=0.
\end{equation*}
The eigenvalues are related to the zeros of the Bessel functions by
\begin{equation*}
 \sigma=j^2_{n{\pm}1,m}+\alpha^2, \;\; {\mbox{and}} \;\; \sigma=j^2_{n,m}+\alpha^2,
\end{equation*}
for $m=0,1,2, \cdots $.
%
%
\section{Damped modes of two-dimensional disturbances}
When the disturbances can be considered to consist of long waves, $\alpha \rightarrow 0$, then the disturbances are predominantly confined within planes normal to the axis of the pipe. In this approximation, the disturbances are taken to be independent of $x$, and thus are proportional to the waves of 
\begin{equation*}
 \exp \big( {-} \omega t + \ri n \theta \big),
\end{equation*}
where $\omega$ is the eigenvalue to be determined. Setting $\alpha=0$ in equation (\ref{eq:pressure}), the bounded solution of the pressure is simply $A_{0}r^n$, where $A_{0}$ is a constant. 

The equations of the motion reduce to
\begin{equation} \label{u-comps-2d}
\begin{split}
 \psix''+\psix'/r
-n^2 \psix/r^2 + \omega R \psix - \ri R V' \psir & =0, \\ 
 \psir''+\psir'/r
 -(n^2+1) \psir/r^2 -2n \psit/r^2 + \omega R \psir & = -\ri Rp', \\
\psit''+\psit'/r
 -(n^2+1) \psit/r^2 -2n \psir/r^2 + \omega R \psit & = \ri Rnp/r, \\
 \psi'_{r}+{\psi_{r}}/r+n{\psi_{\theta}}/r & = 0.
\end{split}
\end{equation}
All the wall boundary conditions remain unchanged.  By analogy, the auxiliary functions are
\begin{equation*}
\psi''_{\pm}+\psi'_{\pm}/r + \Big( \gamma^2-(n \pm 1)^2/r^2 \Big) 
{\psi}_{\pm}=q_{\mp}(r),
\end{equation*}
where $\gamma^2=\omega R$, $q_-=0$, and $q_+=2A_{0} n r^{n-1}$. Thus the bounded solutions are expressed in terms of the Bessel functions
\begin{equation*}
 \gfrac{\psir(r)}{\psit(r)}=A_{+} J_{n+1}(\gamma r) \pm A_{-} J_{n-1}(\gamma r) \pm n A_{0} \int_0^r   U(r,s) s^n \rd s,	
\end{equation*}
where $A$'s are constants, and
\begin{equation*}
 U(r,s)=J_{n-1}(\gamma r) \: Y_{n-1} (\gamma s) - J_{n-1}(\gamma s) \: Y_{n-1} (\gamma r).
\end{equation*}
For non-vanishing $A_{\pm}$, applying the wall boundary conditions yields
\begin{equation*}
 J_{n+1}(\gamma)=0, \;\;\; \mbox{and} \;\;\; \int_0^1 s^n J_{n-1}(\gamma s) \rd s = J_n (\gamma)/\gamma=0.
\end{equation*}
All these disturbances are in the form of stationary waves. The eigen-modes coincide with the zeros of the Bessel functions and are given by
\begin{equation*}
 \begin{split}
	\omega= j^2_{n,m} \:R^{-1}, \;\;\; \mbox{and} \;\;\; \omega = j^2_{n+1,m}\: R^{-1}.
 \end{split}
\end{equation*}
The least damped mode is $ \sim O(n^2/R)$ as $n \rightarrow \infty$ according to (\ref{n-wall-tmodes-Airy}).  
For example, the modes ($\omega R$) for $n=0,1,2$ are $5.7832$, $14.6820$ and $26.3746$ respectively.
%
%
\section{Conclusion}
For every fixed periodicity $n \geq 0$, we have shown that the pipe Poiseuille flow decays exponentially in time with respect to the three infinitesimal disturbances at all wave numbers and Reynolds numbers. There exist denumerable discrete eigen-modes for given $\alpha$ and $R$. The eigen-values all have negative imaginary part - a characteristic closely connected with certain entire functions of the parameters $\alpha$, $R$ and $n$. Some asymptotic relations have been derived and they compare favourably with the full numerical solutions of the linearized pipe flow over a wide range of $n$. For the two-dimensional disturbances $\alpha \rightarrow 0$, the flow is found to be in a stationary stage. 

The present study approximates the stage of linear diffusion due to viscosity in the equations of motion. This linear stage occurs in every flow of fluid motion over a short time from the start of the motion. As we have not specifically given an initial value, we thus interpret the present result as the evolution of a flow with the parabolic profile as its starting distribution.  

It is known that the non-linearity, $(u.\nabla)u$, in the Navier-Stokes equations inherently defines a state of fluid motion known as turbulence (Lam 2013). In fluid mechanics, turbulence is the general solution of the vorticity equation. The dynamic structure of turbulence comprises space-time superposition of multitudinous vorticity eddies, which have mathematical presentations as multiple integral convolutions of initial vorticity and the diffusion or heat kernel. For given initial data, the laminar-turbulent transition is nothing more than an evolutionary process in which vorticity eddies of smaller scales successively emerge in large quantity. The linearization procedure employed in our study has effectively suppressed the non-linearity and hence it cannot touch any essential aspect of turbulence. In brief, the diffusive linear development in pipe flow has no direct connection with the transition process. Application of the linearized equations of motion beyond the diffusion state ought to be erroneous and eventually results in unreliable prediction.
%
%
\vspace{10mm}

\noindent 
25 November 2014

\noindent 
\texttt{f.lam11@yahoo.com}
\vspace{5mm}
\appendix{}
Each of Whittaker's equations, 
\begin{equation} \label{whit}
{\phi''_{\pm}}(z)+\Big(\: -\frac{1}{4}+\frac{\kappa_{\pm}}{z}+\frac{1-(n{\pm}1)^{2}}{4z^{2}} \:\Big) {\phi_{\pm}}(z)={\phi''_{\pm}}+q(z) {\phi_{\pm}}=0, 
\end{equation}
admits two linearly independent solutions, $z=z_0 r^2$,
\begin{equation} \label{whit-func}
\begin{split}
M_{{\kappa_{\pm}},{\mu_{\pm}}}(r)&=z^{b_{\pm}/2} \re ^{-z/2}M(a_{\pm},b_{\pm},z),\\
W_{{\kappa_{\pm}},{\mu_{\pm}}}(r)&=z^{b_{\pm}/2} \re ^{-z/2}U(a_{\pm},b_{\pm},z),
\end{split}
\end{equation}
where $b_+ =n + 1$, $b_-=n$, $\mu_{\pm}=(b_{\pm}-1)/2$, and $a_{\pm}=b_{\pm}/2-\kappa_{\pm}$ (cf. (\ref{kap})).
In particular,
\begin{equation}
 \begin{split}
	\Real q & =-\frac{1}{4}\Big( 1-\frac{1-c_r}{r^2} \Big),\\
	\Imag q & =-\frac{1}{4r^2} \Big( \:c_i+\frac{\alpha}{R} + \frac{1}{\alpha R r^2}\big( (n \pm 1)^2-1 \big) \:\Big).
 \end{split}
\end{equation}
Multiplying the equations by the conjugate function $\phi_{\pm} ^{*}$ and integrating from $0$ to $r$, we have for the real and imaginary parts:
\begin{equation}
\begin{split}
(\sqrt{\alpha R})^{-1} & \: \Real \big[ \phi_{\pm}' \phi_{\pm}^* (r) \;\big]  - \int_0^r   \abs{\phi_{\pm}'}^2 \rd r^2 + \int_0^r  \Imag q \: \abs{\phi_{\pm}}^2 \rd r^2=0, \\
(\sqrt{\alpha R})^{-1} & \:	\Imag \big[ \phi_{\pm}' \phi_{\pm}^* (r) \; \big] - \int_0^r     \Real q \: \abs{\phi_{\pm}}^2 \rd r^2=0. 
\end{split}
\end{equation}
We take $\phi_{\pm}$ as the function $M_{{\kappa_{\pm}},{\mu_{\pm}}}$. If $c_i > -\alpha/R$, $\Imag q < 0$ for $0 < r \leq 1$. In the light of the oscillation theorems in the complex plane (see, for example, Chapter XXI of Ince 1927), neither $M_{{\kappa_{\pm}},{\mu_{\pm}}}$ nor $M'_{{\kappa_{\pm}},{\mu_{\pm}}}$ can have (complex) zeros over any interval $r>0$
because $M_{{\kappa_{\pm}},{\mu_{\pm}}}(r=0)=0$, namely, the zero boundary condition at the pipe centre $r=0$.
Note that $\Imag q$ remains unchanged for $n \rightarrow -n$. Therefore the particular case $n=1$ is equivalent to $n=-1$. Hence, $M_{{\kappa_{\pm}},{\mu_{\pm}}}(z_0) \neq 0$ and $M'_{{\kappa_{\pm}},{\mu_{\pm}}}(z_0) \neq 0$ for $c_i > -\alpha/R$.

Kummer's function, $M(a,b,z)$, is an entire function of $z$, $b$ and $a$ for $a \neq -1,-2,...$, and $U(a,b,z)$ is an analytic function of $z$ in the plane cut along the negative real axis. Dropping the suffix for $a$ and $b$ temporarily for the sake of simplicity, 
for fixed $a$ and $b$, Kummer's functions have the following asymptotic expansions, as $|z| \rightarrow \infty$ (see, for example, Abramowitz \& Stegun 1972; Olver 1997). For $-{\pi}/2 < {\text{arg}}\;z < 3/2{\pi}$,
\begin{equation} \label{eq:m-asymp}
	M(a,b,z)=  h_1 \sum_{k=0}^{\infty}\frac{(a)_{k}(1+a-b)_{k}}{k!}(-z)^{-k} +h_2\sum_{k=0}^{\infty}
	\frac{(b-a)_{k}(1-a)_{k}}{k!} z^{-k},
\end{equation}
where $h_1=e^{\ri {\pi}a}{z^{-a}}{\Gamma(b)}/{\Gamma(b-a)}$, and $h_2=e^{z}z^{a-b} {\Gamma(b)}/{\Gamma(a)}$.
Similarly, 
\begin{equation}	\label{eq:u-asymp}
U(a,b,z)=z^{-a}\sum_{k=0}^{\infty}\frac{(a)_{k}(1+a-b)_{k}}{k!}(-z)^{-k},\;-3{\pi}/2 < {\text{arg}}\;z < 3{\pi}/2. 
\end{equation}
In practice, we have used two different approaches for the evaluation of these functions. The first one is a {\emph{careful}} numerical implementation of the series summations. For moderate to large values of $\alpha R$, it is essential to scale $a$ and $b$ by some suitable norm so that calculation overflows may be prevented. The second one is to numerically integrate Kummer's equation, 
\begin{equation*}
zw''+(b-a)w'-aw=0,
\end{equation*}
in the complex $z$-plane by a fourth-order Runge-Kutta scheme with adaptive step size control.
For the values of $|z| \leq 0.25$, the series summation for $M(a,b,z)$ converges rapidly and hence it provides a convenient initial value for the integration. To calculate $U(a,b,z)$ when $|z|$ are small but bounded away from the singularity, use has been made of the various approximation formulas.
\appendix{}
The conditions for solving (\ref{greens-gn}) can be put in matrix notation, for $k=2,3$,
\begin{gather*}
  A^k_n X^k = B^k, \\
  \begin{align*}
  \mbox{where}\;\;\;X^k&=\{A_1^k \;\;A_3^k \;\;A_5^k \;\;B_1^k\;\;B_2^k\;\;B_3^k\;\;B_4^k\;\;B_5^k\;\;B_6^k \}^T, \\
	B^2&=\{0 \;\; 0 \;\; 0 \;\; 0 \;\; 0 \;\; 0 \;\; 0 \;\; 0 \;\; {-1} \}^T, \\
	B^3&=\{0 \;\; 0 \;\; 0 \;\; 0 \;\; 0 \;\; 0 \;\; 0 \;\; 0 \;\; {-1/t} \}^T. \\
	\end{align*}
\end{gather*}
The determinant of $A^2_n$ is given by
\begin{equation*}
\begin{split}
&
\begin{vmatrix}
 0 & 0 & 0 & \psirone(1)  &  \psirtwo(1) & \psirthree(1)  & \psirfour(1) & \psirfive(1) & \psirsix(1) \\
 0 & 0 & 0 & \proneone(1) &  \proneone(1) & \prthreeone(1) & \prfourone(1) & \prfiveone(1)& \prsixone(1) \\
 0 & 0 & 0 & \psitone(1)  &  \psittwo(1) & \psitthree(1)  & \psitfour(1) & \psitfive(1) & \psitsix(1) \\
 -\psirone  & -\psirthree  & -\psirfive &  \psirone  & \psirtwo & \psirthree  & \psirfour & \psirfive  & \psirsix \\
 -\psixone & -\psixthree & -\psixfive  &  \psixone  & \psixtwo & \psixthree  & \psixfour & \psixfive  & \psixsix \\
 -\psitone & -\psitthree & -\psitfive & \psitone  & \psittwo & \psitthree  & \psitfour & \psitfive  & \psitsix \\
 -\pxoneone & -\pxthreeone & -\pxfiveone &  \pxoneone  & \pxtwoone & \pxthreeone  & \pxfourone & \pxfiveone  & \pxsixone \\
 -\ptoneone & -\ptthreeone & -\ptfiveone &  \ptoneone  & \pttwoone & \ptthreeone  & \ptfourone & \ptfiveone  & \ptsixone \\
 -\pxonetwo & -\pxthreetwo & -\pxfivetwo & \pxonetwo & \pxtwotwo & \pxthreetwo  & \pxfourtwo & \pxfivetwo  & \pxsixtwo \\
\end{vmatrix} 
 \\
 & \\
& =\Delta_n(\alpha,R,n,c) W_{x}. 
\end{split}
\end{equation*}
Similarly, $\abs{A^3_n}$ has the similar form with the only difference that $\psix$'s in the last row are replaced by $\psit$'s, 
\begin{equation*}
\abs{A^3_n}=\Delta_n(\alpha,R,n,c) W_{\theta}.
\end{equation*}
$W_x$ and $W_{\theta}$ denote the $6{\times}6$ determinant at the lower right hand corner of $\abs {A^k_n}$.
The independent variable for the functions in the lower 6 rows in both determinants is $t$. Neither $W_x $ nor $W_{\theta} $ can vanish as they are the Wronskians of the linearly independent solutions of (\ref{odes}). The singularity of Green's function coincides with the dispersion relation. Denote the cofactor of the element $a^k_{ij}$ of $\abs{A^k_n}$ by $\varLambda^k_{ij}$, and $G_{lk}(r,t)$ by $g_k(r,t;c=0),l=1,2,3$. By Cramer's rule, the components are found to be
\begin{equation*}
G_{12}(r,t)= \frac{1}{d_2}
\begin{cases}
 \varLambda^2_{91}(t) {\psir}_1(r) + \varLambda^2_{92}(t) {\psir}_3(r) + \varLambda^2_{93}(t) {\psir}_5(r)  \;\;\; \hspace{0.35cm}\mbox{for}\;r{\leq}t, \\
 & \\
 \varLambda^2_{94}(r) {\psir}_1(t) + \varLambda^2_{95}(r) {\psir}_2(t) + \varLambda^2_{96}(r) {\psir}_3(t) \;+ \\
 \;\;\; \varLambda^2_{97}(r) {\psir}_4(t) + \varLambda^2_{98}(r) {\psir}_5(t) + \varLambda^2_{99}(r) {\psir}_6(t) \hspace{0.35cm} \mbox{for}\;r{\geq}t, 
\end{cases}
\end{equation*}
To simplify the writings, we introduce the notation $G_{12}(r,t)=G_{12}(r,t; d_2,\varLambda^2, \psir)$.
Then the remaining components are given by 
\begin{equation*} 
\begin{split}
G_{22}(r,t)&=G_{22}(r,t; d_2,\varLambda^2, \psix),\;\;\;G_{32}(r,t)=G_{32}(r,t; d_2,\varLambda^2, \psit), \\
G_{13}(r,t)&=G_{13}(r,t; d_3,\varLambda^3, \psir), \;\;\;G_{23}(r,t)=G_{23}(r,t; d_3,\varLambda^3, \psix), \\
G_{33}(r,t)&=G_{33}(r,t; d_3,\varLambda^3, \psit), \\
\end{split}
\end{equation*}
where $d_2=- \; W_x \; A^2_n(\alpha,R,n,c=0)$, and $d_3=-n\;W_{\theta} \; A^3_n(\alpha,R,n,c=0)$.
We arrive at a system of integral equations
\begin{equation} \label{eq:sys-intgleqns}
\begin{split}
{\psir}(r) & = {\lambda}\int_{0}^{1}   N_{11} \psir(t) \rd t+{\lambda}\int_{0}^{1}   N_{12} \psix(t) \rd t+{\lambda}\int_{0}^{1}   N_{13} \psit(t) \rd t, \\
\psix(r) & = {\lambda}\int_{0}^{1}    N_{21}\psir(t) \rd t+{\lambda}\int_{0}^{1}   N_{22} \psix(t) \rd t+{\lambda}\int_{0}^{1}   N_{23} \psit(t) \rd t, \\
\psit(r) & = {\lambda}\int_{0}^{1}    N_{31}\psir(t) \rd t+{\lambda}\int_{0}^{1}   N_{32} \psix(t) \rd t+{\lambda}\int_{0}^{1}   N_{33} \psit(t) \rd t, \\
\end{split}
\end{equation}
where $\lambda=- \ri \alpha R c$, and the kernels are given by
\begin{equation*}
\begin{split}
N_{11}& = \alpha G_{12} + n G_{13},\; N_{12} = - {\partial G_{12} }/{\partial t }, \; N_{13} = G_{13}- {\partial (tG_{13}) }/{\partial t },\\
N_{21}& = \alpha G_{22} + n G_{23},\; N_{22} = - {\partial G_{22} }/{\partial t }, \; N_{23} = G_{23}- {\partial (tG_{23}) }/{\partial t },\\
N_{31}& = \alpha G_{32} + n G_{33},\; N_{32} = - {\partial G_{32} }/{\partial t }, \; N_{33} = G_{33}- {\partial (tG_{33}) }/{\partial t }.\\
\end{split}
\end{equation*}
%
%
%

\label{lastpage}
\end{document}